\documentclass[preprint,12pt,sort&compress]{elsarticle}

\usepackage{amsmath,amssymb,mathrsfs}
\usepackage[]{graphicx}
\usepackage[latin1]{inputenc}
\usepackage{color}
\usepackage{enumerate}
\usepackage{hyperref}
\usepackage{caption}
\hypersetup{colorlinks=true,citecolor=blue}

\begin{document}

\begin{frontmatter}

\title{Fusion of laser diffraction and chord length distribution data for estimation of particle size distribution using  multi-objective optimisation}

\author[a1]{Okpeafoh S. Agimelen\corref{cor1}}
\ead{okpeafoh.agimelen@strath.ac.uk}

\author[a1]{Carla Ferreira}

\author[a2]{Bilal Ahmed}

\author[a3]{Javier Cardona}

\author[a1]{Yi-chieh Chen}

\author[a2]{Alastair J. Florence}

\author[a3]{Ivan Andonovic}

\author[a1]{Jan Sefcik}

\author[a4]{Anthony J. Mulholland\corref{cor1}}
\ead{anthony.mulholland@strath.ac.uk}

\cortext[cor1]{Corresponding authors}
\address[a1]{EPSRC Future Manufacturing Hub for Continuous Manufacturing and Advanced Crystallisation, Department of Chemical and Process Engineering, University of Strathclyde, James Weir Building, 75 Montrose Street, Glasgow, G1 1XJ, United Kingdom.}

\address[a2]{EPSRC Future Manufacturing Hub for Continuous Manufacturing and Advanced Crystallisation, Strathclyde Institute of Pharmacy and Biomedical Sciences, Technology and Innovation Centre, University of Strathclyde, 99 George Street, Glasgow, G1 1RD, United Kingdom}

\address[a3]{Centre for Intelligent Dynamic Communications, Department of Electronic and Electrical Engineering, University Of Strathclyde, Royal College Building, 204 George Street, Glasgow, G1 1XW, 
	United Kingdom.
}

\address[a4]{Department of Mathematics and Statistics, University of Strathclyde, Livingstone Tower, 26 Richmond Street, Glasgow, G1 1XH, United Kingdom.}

\begin{abstract}
	
	The in situ measurement of the particle size distribution (PSD) of a suspension of particles presents huge challenges. Various effects from the process could introduce noise to the data from which the PSD is estimated. This in turn could lead to the occurrence of artificial peaks in the estimated PSD. Limitations in the models used in the PSD estimation could also lead to the occurrence of these artificial peaks. This could pose a significant challenge to in situ monitoring of particulate processes, as there will be no independent estimate of the PSD to allow a discrimination of the artificial peaks to be carried out.
	
	Here, we present an algorithm which is capable of discriminating between artificial and true peaks in PSD estimates based on fusion of multiple data streams. In this case, chord length distribution and laser diffraction data have been used. The data fusion is done by means of multi-objective optimisation using the weighted sum approach. The algorithm is applied to two different particle suspensions. The estimated PSDs from the algorithm are compared with offline estimates of PSD from the Malvern Mastersizer and Morphologi G3. The results show that the algorithm is capable of eliminating an artificial peak in a PSD estimate when this artificial peak is sufficiently displaced from the true peak. However, when the artificial peak is too close to the true peak, it is only suppressed but not completely eliminated. 
	
\end{abstract}


\end{frontmatter}

\section{Introduction}
\label{sec1}

The particle size distribution (PSD) is a key quality attribute of the particles in powders. This is particularly important in the manufacturing of particulate products in the pharmaceutical, food, personal care and fine chemicals industries where the success or failure of the process depends heavily on the PSD of the particles \cite{Nagy2013}. In addition, the perfomance of the final products depend heavily on the PSD of the constituent particles \cite{Merkus2009}. 

Various methods and techniques \cite{Washington1992} exist for the estimation of the PSD of a suspension of particles. These methods typically involve the processing of sensor data or analysis of images.
However, each method is limited by the range of applicability of the sensor used in the measurement. For example, imaging methods could yield inaccurate results 
in very dense suspensions where there are too many overlapping objects in the images or insufficient contrast between the objects and image background \cite{Larsen2006,Agimelen2017n2}. Methods based on analysis of laser diffraction data could produce artificial particle size modes if there is multiple scattering or the model used in the analysis is inadequate for the shape of the particles in the suspension \cite{Omer1990,Agimelen2017}. Methods based on analysis of chord length distribution (CLD) could overestimate the fines in a suspension for example, when bubbles are produced in a process and they contribute to the CLD counts \cite{Agimelen2017n2}.

Some of the inaccuracies involved in the estimation of PSD from a single sensor data could be alleviated by adopting the approach of multi-sensor data fusion. This data fusion approach involves integrating data from multiple sensors for greater accuracy of estimated quantities \cite{Khaleghi2013}. Various techniques exist for the implementation of multi-sensor data fusion. They are mostly aimed at dealing with noise and imperfections of data from multiple sensors \cite{Khaleghi2013}.

Here, we adopt the weighted sum approach of multi-objective optimisation  \cite{Deb2001}. This approach is easy to implement as it allows a decision maker to unambiguously choose from a range of mathematically feasible solutions. We demonstrate the applicability of this approach in the estimation of PSD in particulate processes by applying this weighted sum method to combined CLD and laser diffraction data. We vary the weights on the combined data streams to produce a Pareto curve. We then apply a method for estimating the knee point of the Pareto curve at which the desired PSD is chosen. Our results show that some of the artefacts encountered in estimating PSD from single sensors can be eliminated or mitigated by this multi-objective approach.

\section{Materials and experimental procedure}
\label{sec2}

The materials used in this work consisted of polystyrene (PS) particles made by suspension polymerisation \cite{EmidioPhD}. The particles were in a wide range of sizes, hence sieving was carried out to obtain particles in different size ranges. The sieve fraction PS 300-500$\mu$m exhibited multiple peaks upon analysis, and hence was included in this paper as it demonstrated the phenomenon which is the subject of this paper. The particles in the PS 300-500$\mu$m size range were mostly spherical in shape  (see supplementary information for sample images).

Sieve fractions of metformin hydrochloride (MET), which was purchased from Molekula, were also prepared for measurements in this work. The sieve fraction MET 180-250$\mu$m was used in this work. Similar to the PS 300-500$\mu$m sieve fraction, the MET 180-250$\mu$m sieve fraction also exhibited multiple peaks, and hence was included in this paper. The MET particles were rod-like (see supplementary information for sample images), so that the larger sieve fractions contained a significant quantity of particles larger than 1mm. These particles would be too large for the measurement instruments used in this work, hence the larger sieve fractions were not used. Similarly, the smaller sieve fractions contained a significant quantity of fines which would complicate the analysis, so that the smaller sieve fractions were not used. 

A sample consisting of 6.160g of the PS 300-500$\mu$m was suspended in 20.045g of water. The suspension was stirred with an overhead stirrer at 400 rpm. The suspension was monitored with both the focus beam reflectance measurement (FBRM) G400 and particle vision and measurement (PVM) V819 probes from Mettler Toledo. Another sample of the PS 300-500$\mu$m was suspended in water in the Malvern Mastersizer instrument, and subsequently measured. Both the estimated volume based PSD and raw measured scattering intensity data were obtained from the instrument software. Finally, a sample of the PS 300-500$\mu$m sieve fraction was dispersed in the Malvern Morphologi G3 instrument for PSD estimation.

A similar procedure was applied for the MET 180-250$\mu$m sieve fraction. In the case of MET 180-250$\mu$m sieve fraction, a mass of 6.968g of the sample was suspended in 68.902g of isopropanol (IPA) for FBRM and PVM measurement. Samples of the MET 180-250$\mu$m sieve fraction were also analysed with the Malvern Mastersizer and Morphologi instruments as in the case of the PS 300-500$\mu$m sieve fraction.

\section{Method}
\label{sec3}

The fusion of CLD and laser diffraction data was done in a multi-objective optimisation manner in this work. The weighted sum approach \cite{Das1997,Deb2001,Kim2005} in multi-objective optimisation was adopted because of its ease of implementation and interpretation. The procedure for performing the data fusion is described in the following subsections. 

\subsection{Multi-objective optimisation problem}
\label{ssec3-1}

In the weighted sum approach involving two objective functions, the optimisation problem is formulated as follows \cite{Deb2001}
\begin{align}
\underset{\mathbf{X}}\min  \quad F(\mathbf{X,\gamma}) &  = \gamma f_1(\mathbf{X}) + (1-\gamma) f_2(\mathbf{X})\nonumber\\
\textrm{subject to}  \quad \mathbf{X} & \geq \mathbf{0} \nonumber\\
\quad \gamma & \in[0,1],
\label{eq1}
\end{align}
where $\gamma$ is a weight parameter and $F$ is the single objective function formed from the combination of $f_1$ and $f_2$. Each objective function $f_1$ and $f_2$ depends on the $N$-dimensional PSD $\mathbf{X}$. The objective functions $f_i$, $i=1,2$ are such that $f_i: \mathbb{R}^N\mapsto\mathbb{R}$.

In this work each of the objective functions is the sum of squares error between the experimentally measured quantity and the corresponding modelled quantity. Each objective function is weighted by the estimated standard deviation of each measurement \cite{Aster2013,Pedersen1997}. In the case of CLD the objective function is given as
\begin{equation}
f_1 = \sum_{i=1}^{M_1}{\left[\frac{C_i^* - C_i}{\sigma_{1i}} \right]^2},
\label{eq2}
\end{equation}
where $C_i^*$ is the measured CLD and $C_i$ is the modelled CLD. The CLD was modelled using an ellipsoidal CLD model as in previous works \cite{Li2005n1,Agimelen2015,Agimelen2016}. The standard deviation $\sigma_{1i}$ is estimated from multiple CLD measurements. The PSD is discretised into $N=200$ geometrically spaced bins and the CLD into $M_1=100$ geometrically spaced bins \cite{Li2005n1,Agimelen2015,Agimelen2016}.

The elliptical model used in this work \cite{Li2005n1} admits aspect ratios ($r$) within the range $r\in (0,1]$. The aspect ratio is the ratio of the minor to major axes lengths of the ellipse used to model the particle. In this work, the aspect ratio is fixed at $r=1$, implying that a spherical model for the CLD was used. This is to make it consistent with the spherical model used for laser diffraction in this work.

Similar to the case of $f_1$, the objective function $f_2$ for laser diffraction is given as 
\begin{equation}
f_2 = \sum_{i=1}^{M_2}{\left[\frac{I_i^* - I_i}{\sigma_{2i}} \right]^2},
\label{eq3}
\end{equation}
where $I_i^*$ is the experimentally measured scattering intensity by laser diffraction. The modelled scattering intensity $I_i$ is obtained using the Mie model for spherical particles \cite{Bohren1983,Agimelen2017}. This is because of its common use in the software of commercial laser diffraction instruments for particle sizing. The scattering intensity data from the red laser (used in this work) of the Malver Mastersizer instrument is collected in $M_2=47$ bins. These detectors cover the focal plane and wide angle directions \cite{Malvern2015}. Like the case of CLD, the standard deviation $\sigma_{2i}$ is estimated from multiple scattering intensity measurements.

The PSD $X_i$ is modelled as a mixture of $N_b$ log-normal distribution (which is applicable in modelling a large number of physical processes \cite{Crow1988}) basis functions as
\begin{equation}
X_i = \sum_{j=1}^{N_b}{w_j\frac{1}{D_i\sigma_j\sqrt{2\pi}}e^{-\frac{\left(\ln(D_i)-\mu_j\right)^2}{2\sigma_j^2}}}.
\label{eq4}
\end{equation}
The discretised particle size $D_i$ of bin $i$, which is the spherical equivalent diameter, is the geometric mean of the particle sizes in the boundaries of bins $i$ and $i+1$ consistent with the PSD \cite{Agimelen2015,Agimelen2016,Agimelen2017}. The parameters $\mu_j$ and $\sigma_j$ set the position of the peak and the width of each of the log-normal basis function. A minimal number of basis function that allows a good fit to be obtained for each measured quantity is used in the calculations. The weights $w_j$ on the basis functions satisfy the following requirement
\begin{equation}
\sum_j^{N_b}w_j = 1.
\label{eq5}
\end{equation}
This implies that the PSD $X_i$ is parametrised by $w,\mu$ and $\sigma$. Hence the multi-objective optimisation problem in Eq. \eqref{eq1} can be rewritten as 
\begin{align}
\underset{\mathbf{X}(w,\mu,\sigma)}\min \quad F(\mathbf{X}(w,\mu,\sigma),\gamma) & = \gamma f_1(\mathbf{X}(w,\mu,\sigma)) + (1-\gamma) f_2(\mathbf{X}(w,\mu,\sigma))\nonumber\\
\textrm{subject to} \quad \mathbf{X}(w,\mu,\sigma) & \geq \mathbf{0} \nonumber\\
\quad \gamma & \in[0,1] \nonumber\\
\quad \sum_j^{N_b}w_j & = 1 \nonumber\\
\quad \sigma^2 + \mu^2 & \leq a^2.
\label{eq6}
\end{align}
The inequality for $\mu$ and $\sigma$ in Eq. \eqref{eq6} constrains the search for $\mu$ and $\sigma$ to a circular region of specified radius $a$.

The combined objective function $F$ in Eq. \eqref{eq6} is minimised using a gradient based constrained optimisation solver in MATLAB to generate the Pareto optimal solutions. The set of Pareto optimal solutions is obtained using the principle of non-dominance \cite{Deb2001}. 

A solution vector $\mathbf{X}_1$ is said to dominate another solution vector $\mathbf{X}_2$ if these two conditions hold :

\begin{enumerate}
\item The solution $\mathbf{X}_1$ is no worse than $\mathbf{X}_2$ in all objectives. That is, $f_i(\mathbf{X}_1) \leq f_i(\mathbf{X}_2)$, $i=1,2$.
\item The solution $\mathbf{X}_1$ is strictly better than $\mathbf{X}_2$ in at least one objective. That is, $f_i(\mathbf{X}_1) < f_i(\mathbf{X}_2)$ for at least one $i=1,2$.
\end{enumerate}
The set of non-dominated solutions make up the Pareto-optimal set. All solutions in the Pareto-optimal set are mathematically feasible.

As the objective functions $f_1$ and $f_2$ in Eqs. \eqref{eq2} and \eqref{eq3} occur at different scales, a normalisation is carried out to put them on the same scale. This is done using the Utopian $f_i^U$ and Nadir $f_i^N$ points estimated for each function \cite{Deb2001,Kim2005} as
\begin{equation}
\tilde{f}_i = \frac{f_i - f_i^U}{f_i^N - f_i^U}, \quad i=1,2,
\label{eq7}
\end{equation}
where the Utopian points are the set of global minima for each objective function when minimised separately, and the Nadir points are the set of upper bounds of each objective function in the Pareto set \cite{Deb2001}.

The multi-objective optimisation problem in Eq. \eqref{eq6} is then solved by minimising the combined objective function $F$, where each of the constituent objective function $f_1$ and $f_2$ has been replaced by it corresponding normalised version using Eq. \eqref{eq7}.

For each value of the weight parameter $\gamma$ in Eq. \eqref{eq6}, it may be possible to find a non-dominated solution, and corresponding values of $\tilde{f}_1$ and $\tilde{f}_2$. Hence, a vector of $\gamma$ values can be constructed to obtain a possible set of non-dominated solutions, which correspond to different pairs of $\tilde{f}_1$ and $\tilde{f}_2$ values. The graph of these $\tilde{f}_1$ and $\tilde{f}_2$ values constitute the Pareto curve.

The initially estimated Pareto curve obtained by the above procedure typically contain non-uniformly distributed points even for uniformly spaced values of $\gamma$. This is a common issue \cite{Das1997,Deb2001,Kim2005} with the weighted sum approach described in Eq. \ref{eq6}. However, the weight parameter $\gamma$ in Eq. \eqref{eq6} is related to the slope of the Pareto curve as \cite{Das1997}
\begin{equation}
\gamma = \frac{1}{1 - \frac{\partial \tilde{f}_1}{\partial \tilde{f}_2}}.
\label{eq8}
\end{equation}
This assumes a functional form of the Pareto curve say $\psi$ so that the mapping $\psi \colon \tilde{f}_2 \mapsto \tilde{f}_1$ holds, that is $\tilde{f}_1=\psi(\tilde{f}_2)$ \cite{Das1997}. This allows a new vector of weights $\gamma$ to be estimated that gives a Pareto curve with better spread of points. The steps to estimate the Pareto curve are outlined in the next section.

\subsection{Multi-objective optimisation algorithm}
\label{ssec3-2}

\subsubsection{Estimating guess starting points for searching}
\label{sssec3-2-1}

The process of minimising the objective functions $F$ in Eq. \eqref{eq6}, $f_1$ in Eq. \eqref{eq2} and $f_2$ in Eq. \eqref{eq3} involves searching for $w_j$, $\mu_j$ and $\sigma_j$ values in Eq. \eqref{eq4} for a PSD, such that the corresponding CLD and/or scattering intensity data give good fits to the corresponding experimentally measured data. However, as the gradient based algorithms used can only guarantee finding values at the local minima of corresponding function, then it is necessary to start the search with values of the quantities such that the corresponding objective functions are close to their global minima. This is done using the method of truncated singular value decomposition (TSVD) \cite{Aster2013} as outlined below.

\begin{enumerate}
\item Construct transformation matrices $\mathbf{A}_1$ and $\mathbf{A}_2$ with $N$ columns. The columns of $\mathbf{A}_1$ are the simulated CLDs $\mathbf{C}$ for different particle sizes $D_1$ to $D_N$. Similarly, the columns of $\mathbf{A}_2$ are the simulated scattering intensities $\mathbf{I}$ for different particle sizes as in the case of $\mathbf{A}_1$.
\item Obtain the singular values $s_{ki}, i=1,2, ..., N_{ks}, k=1,2$ of matrices $\mathbf{A}_k, k=1,2$. The singular values are ordered such that $s_{k1}<s_{k2}< ... <s_{kN_{ks}}, k=1,2$.
\item Obtain the PSDs $\mathbf{X}_{1i} = \tilde{\mathbf{A}}_{1i}\mathbf{C}, i=1,2, ..., N_{1s}-1$ and $\mathbf{X}_{2i} = \tilde{\mathbf{A}}_{2i}\mathbf{I}, i=1,2, ..., N_{2s}-1$ by direct matrix multiplication, where $\tilde{\mathbf{A}}_k, k=1,2$ are the pseudo inverse of matrices $\mathbf{A}_k,k=1,2$. Each of the pseudo inverse is constructed with a tolerance on the singular values, such that, only singular values $s > s_{ki}, i=1,2, ..., N_{1s}-1, k=1,2$ are used. 
\item Since the method of TSVD does not guarantee non-negative solutions, take the absolute values of the elements of each of the PSD vectors $\mathbf{\tilde{X}}_{ki}, i=1,2, ..., N_{1s}-1, k=1,2$.
\item For each PSD $\mathbf{\tilde{X}}_{ki}, i=1,2, ..., N_{1s}-1, k=1,2$, fit the PSD $\mathbf{X}$ in Eq. \eqref{eq4} for a specified value of $N_b$ to obtain different sets of values of $w_{ki}$, $\mu_{ki}$ and $\sigma_{ki}$.
\item For each set of values of $w_{ki}$, $\mu_{ki}$ and $\sigma_{ki}$, obtain the $L_2$ norm $L = \|\mathbf{\tilde{X}}_{ki} - \mathbf{X}\|, i=1,2, ..., N_{ks}-1, k=1,2$.
\item Choose the set of $w_k$, $\mu_k$ and $\sigma_k,k=1,2$ with the smallest value of $L$ as the guess starting points for a subsequent search algorithm.
\end{enumerate}

\subsubsection{Estimating the Nadir and Utopian points}
\label{sssec3-2-2}

In order to perform the normalisation in Eq. \eqref{eq7}, it is necessary to estimate the Nadir and Utopian points of the objective functions $f_1$ and $f_2$  in Eq. \eqref{eq6} as follows.

For the Nadir points:
\begin{enumerate}
\item Set $N_b=1$ in Eq. \eqref{eq4}.
\item Obtain guess starting values for $\mu$ and $\sigma$ in Eq. \eqref{eq4} for each objective function $f_1$ and $f_2$ using the method of TSVD outlined in section \ref{sssec3-2-1}.
\item Using the respective starting values for $\mu$ and $\sigma$, minimise each of the objective function $f_1$ and $f_2$ separately. Obtain the corresponding PSDs $\mathbf{X}_1^0$ and $\mathbf{X}_2^0$ at the minimum of each function.
\item Estimate the Nadir point as $f_i^N = \max[f_i(\mathbf{X}_1^o), f_i(\mathbf{X}_2^o)], i=1,2$. This is similar to the approach used in \cite{Kim2005}.
\end{enumerate}

For the Utopian points:
\begin{enumerate}
\item Set $N_b=1$ in Eq. \eqref{eq1}, set tolerances $Tol_1$ and $Tol_2$.
\label{step21}
\item Obtain guess starting values for $\mu$ and $\sigma$ in Eq. \eqref{eq4} for objective function $f_1$ using the method of TSVD.
\label{step22}
\item Using the starting values for $\mu$ and $\sigma$, obtain the minimum $f_{11}$ of the objective function $f_1$.
\label{step23}
\item $\begin{array}[t]{l}
\textrm{set counter}\, i = 2 \\
\textrm{enter while loop:} \\
\textrm{set}\, N_b = N_b + 1 \\
\textrm{Repeat steps \ref{step22} and \ref{step23} to obtain}\, f_{1i}\, \textrm{and}\, w_i,\, \mu_i,\, \sigma_i\, \textrm{in}\, \textrm{Eq.}\,\eqref{eq4}.\\
\textrm{if(abs}(f_{1i} - f_{1i-1}) \leq Tol_1) \\
\textrm{Set}\, N_{b1} = N_b, \, f_1^U = f_{1i} \\
\textrm{accept}\, w_i, \mu_i\, \textrm{and}\, \sigma_i\, \textrm{values} \\
\textrm{exit while loop} \\
\textrm{else} \\
\textrm{set}\, i = i + 1 \\
\textrm{end while loop}
\end{array}$
\label{step24}
\item Repeat steps \ref{step22} to \ref{step24} for $f_2$ using $Tol_2$ and obtain $f_2^U, N_{b2}, w, \mu$ and $\sigma$ values for $f_2$.
\end{enumerate}

\subsubsection{Obtaining combined solution}
\label{sssec3-2-3}

To obtain the combined solution from the two objective functions $\tilde{f}_1$ and $\tilde{f}_2$, minimise the combined objective function $F$ in Eq. \eqref{eq6} (with $f_1$ and $f_2$ replaced by $\tilde{f}_1$ and $\tilde{f}_2$ respectively) as follows.

\begin{enumerate}
\item Set $N_b = max[N_{b1}, N_{b2}]$.
\label{step31}
\item If $N_{b1}\geq N_{b2}$, choose the starting values for $w$, $\mu$ and $\sigma$ corresponding to $f_1$ obtained in step \ref{step24} in section \ref{sssec3-2-2}. Else, choose the values corresponding to $f_2$.
\label{step32}
\item Create a vector of $\gamma$ values in Eq. \eqref{eq6}.
\label{step33}
\item For each value of $\gamma$ in the vector of $\gamma$ values in step \ref{step33}, minimise the objective function $F$ in Eq. \eqref{eq6}.
\label{step34}
\item Apply the principle of non-dominance described in section \ref{ssec3-1} to obtain the Pareto optimal set corresponding to the initial vector of $\gamma$ values. Delete all dominated solutions and corresponding $\gamma$ values.
\label{step35}
\end{enumerate}

\subsubsection{Obtaining better spread of points on the Pareto curve}
\label{sssec3-2-4}

After obtaining an estimate of the Pareto curve using the procedure described in section \ref{sssec3-2-3}, then apply the following procedure to get another estimate of the Pareto curve with more uniformly spaces points.

\begin{enumerate}
\item Let the functional form of the mapping $\psi: \tilde{f}_2 \mapsto \tilde{f}_1$ be given as $\tilde{f}_1 = \alpha_1\exp(\tilde{f}_2) + \alpha_2\exp(\tilde{f}_2)$, 
$\alpha_1$ and $\alpha_2$ are arbitrary fitting parameters. The form of mapping chosen in this step depends on the shape of the Pareto curve estimated in section \ref{sssec3-2-3}. Fit the curve $\psi$ to the initially estimated Pareto curve to obtain values of the parameters $\alpha_1$ and $\alpha_2$.
\label{step41}
\item Compute the Euclidean distance $d$ between successive points on the initially estimated Pareto curve. Set the constant spacing $\tilde{d}$ between successive points on the new Pareto curve to be estimated as the minimum value of $d$.
\label{step42}
\item Find the $x$ coordinate of the point of intersection of a circle of radius $\tilde{d}$ centred on the first point $(\tilde{f}_{21},\tilde{f}_{11})$ on the initially estimated Pareto curve and the straight line joining the first two points. This $x$ coordinate corresponds to the value $\tilde{f}_{22}$ of the objective function $\tilde{f}_2$, such that the arc length between $(\tilde{f}_{21},\tilde{f}_{11})$ and $(\tilde{f}_{22},\tilde{f}_{12})$ is approximately equal to $\tilde{d}$, where $\tilde{f}_{12}$ is the $y$ coordinate of the point whose $x$ coordinate is $\tilde{f}_{22}$.
\label{step43}
\item Estimate the value of $\tilde{f}_{12}$ by substituting $\tilde{f}_{22}$ for $\tilde{f}_2$ in the mapping $\psi$ in step \ref{step41}.
\label{step44}
\item Repeat steps \ref{step43} and \ref{step44} to get successive points on the fitted curve $\psi$ whose arc length distance is approximately eaqual to $\tilde{d}$. This gives a discretisation of the fitted curve $\psi$ at the points $(\tilde{f}_{2i},\tilde{f}_{1i}), i=1,2,...,N_p$ such that the arc length distance between successive points is $\tilde{d}$, and the number of these points is $N_p$.
\label{step45}
\item Compute  the derivative given in Eq. \eqref{eq8} at the points $(\tilde{f}_{2i},\tilde{f}_{1i}), i=1,2,...,N_p$ to obtain a new $\gamma$ vector.
\label{step46}
\item Minimise the objective function $F$ in Eq. \eqref{eq6} again using this new $\gamma$ vector, and obtain an improved Pareto curve with a better spread of points.
\label{step47}
\end{enumerate}

\subsubsection{Choosing a solution on the Pareto curve}
\label{sssec3-2-5}

Since all solutions on the Pareto curve are mathematically feasible, the solution chosen is just a decision making process depending on how much weight the decision maker chooses to place on a particular sensor data stream. However, to have a fully automated algorithm, it is necessary to have a criterion for choosing a final solution, independent of the decision maker. The solution at the knee of the Pareto curve is chosen for this purpose here. The boundary line method \cite{Deb2011} for estimating the knee point of the Pareto curve is applied here. This involves drawing a boundary line between two extreme points in the Pareto curve. Then the perpendicular distance of each point on the curve from the boundary line is calculated. The point with the maximum perpendicular distance is chosen as the knee point \cite{Deb2011}.

\section{Results and discussions}
\label{sec4}

As mentioned in the introductory section, the data fusion approach developed in this work allows a discrimination between artificial and real PSD peaks. The artificial peaks are either completely eliminated, when they are sufficiently separated from the true peak, or they get suppressed. These cases are illustrated with results from the two samples analysed in this work. 
The case where an artificial peak is suppressed is illustrated with results from the PS 300-500$\mu$m sample. 
The MET 125-180$\mu$m sample shows an example where an artificial peak (from one sensor modality) that is sufficiently separated from the true peak emerges, and hence gets eliminated. Alongside, it shows an artificial peak from the other sensor modality which is not sufficiently spaced from the true peak, which only gets suppressed. These results are summarised below.

\begin{figure}[htb]
\centerline{\includegraphics[width=\textwidth]{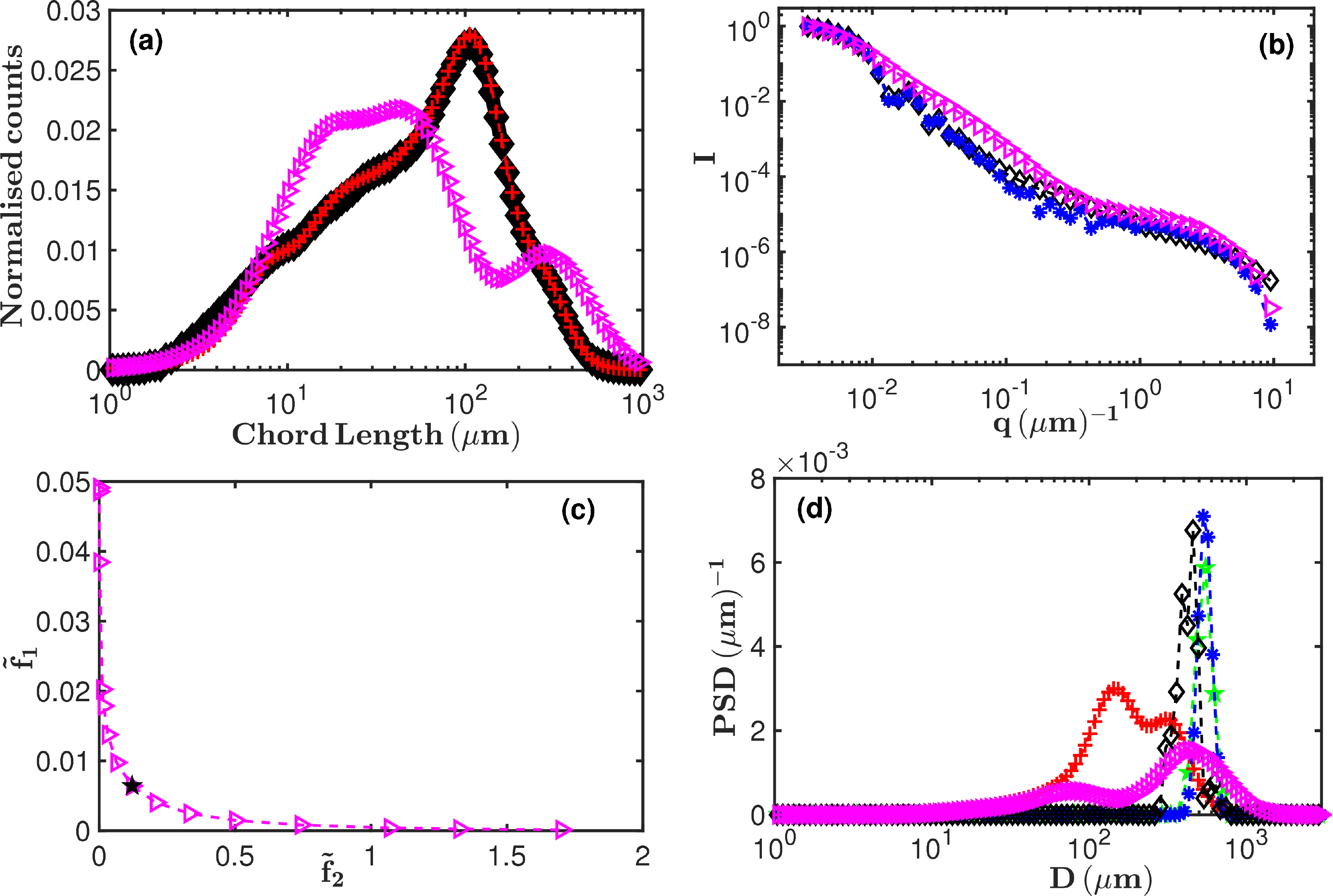}}
\caption{The measured (black diamonds) and estimated CLDs from the objective function $f_1$ in Eq. \eqref{eq2} (red crosses) and at the knee point of the Pareto curve (magenta triangles) for the PS 300-500$\mu$m sample are shown in (a). 
	The measured (black diamonds) and estimated scattering intensities from the objective function $f_2$ in Eq. \eqref{eq3} (blue filled symbols) and the knee point of the Pareto curve (magenta triangles) for the PS 300-500$\mu$m sample are shown in (b). The Pareto curve obtained from the bi-objective function $F$ in Eq. \eqref{eq6} for the PS 300-500$\mu$m sample are shown by the magenta diamonds in (c). The knee point is indicated by the black filled symbol. Volume based PSDs estimated by various methods: black diamonds (Morphologi), green pentagrams (Mastersizer), red crosses (objective function $f_1$), blue filled symbols (objective function $f_2$) and magenta diamonds (at the knee point of the Pareto curve in (c)) for the  PS 300-500$\mu$m sample are shown in (d).}
\label{fig1}
\end{figure}

The estimated volume based PSD obtained from the Morphologi instrument\footnote{The estimated PSDs from the Morphologi instrument is used as reference in this work because the materials had already been manufactured by various processes and sieved. Samples of the materials were then suspended for inline sensor measurement as well as offline measurement with the Morphologi instrument. Since this did not involve any particle recovery from suspension by means of filtering, drying and subsequent dispersion in the Morphologi instrument, the estimated PSD with the Morphologi instrument, which is a highly focused imaging instrument, will be a good representation of the sizes of the particles in the materials.} for the PS 300-500$\mu$m sample suggests a single particle size mode at $D\approx450\mu$m as shown by the black diamonds in Fig. \ref{fig1}(d). This agrees with the estimated volume based PSD by the Mastersizer for the same sample with a single particle size mode at $D\approx500\mu$m as shown by the green pentagrams in Fig. \ref{fig1}(d). This estimate from the Mastersizer coincides with that from the minimisation of objective function $f_2$ in Eq. \ref{eq3} for laser diffraction, as shown by the blue filled symbols in Fig. \ref{fig1}(d). 

However, the estimated volume based PSD (red crosses in Fig. \ref{fig1}(d)) obtained from the minimisation of objective function $f_1$ in Eq. \eqref{eq2} for CLD shows a mode at $D\approx150\mu$m alonside that at $d\approx300\mu$m. the particle size mode at $D\approx150\mu$m is most likely an artifact judging by the estimate from the Morphologi instrument. The reason for this could be due to the bumps on the surface of the PS 300-500$\mu$m particles (see supplementary information for sample images), which results to significant chord splitting, and hence an artificial high counts of short chords. This introduces a left shoulder on the measured CLD as shown by the black diamonds in Fig. \ref{fig1}(a). This causes the optimisation solver to introduce another particle size mode at $D\approx150\mu$m in order to fit the experimentally measured CLD as shown by the red crosses in Fig. \ref{fig1}(a).

The Pareto curve obtained from the minimisation of the combined objective function $F$ in Eq. \eqref{eq6} for the PS 300-500$\mu$m is shown by the magenta triangles in Fig. \ref{fig1}(c), and the knee point is indicated by the black filled symbol. The estimated volume based PSD at this knee point, which is shown by the magenta triangles in Fig. \ref{fig1}(d), shows that the artificial peak at $D\approx150\mu$m associated with the estimated volume based PSD from CLD has been suppressed. The main peak of this estimated volume based PSD at the knee point of the Pareto curve is close to that of the estimated volume based PSD from the offline Morphologi. Hence, the estimated volume based PSD obtained at the knee point of the Pareto curve is a better representation of the particle sizes in the PS 300-500$\mu$m material than the corresponding estimate from the CLD. The estimated volume based PSDs from the Matersizer and that obtained from the minimsation of the objective function $f_2$ are also good representation of the sizes of the particles in the PS 300-500$\mu$m materials in this case. This is due to their closeness to the corresponding estimate from the offline Morphologi instrument and the absence of any artificial peak.

Eventhough the estimated volume based PSD at the knee point of the Pareto curve is a good representation of the PS 300-500$\mu$m material, the estimated CLD corresponding to this PSD at the knee point shows a poor fit to the experimentally measured CLD, as shown by the magenta triangles in Fig. \ref{fig1}(a). Similarly, the estimated scattering intensity at the knee point of the Pareto curve (magenta triangles in Fig. \ref{fig1}(b)) shows a poor fit to the experimentally measured scattering intensity (black diamonds in Fig. \ref{fig1}(b)). However, the estimated scattering intensity (blue filled symbols in Fig. \ref{fig1}(b)) obtained by minimising the objective function $f_2$ for laser diffraction, shows a better fit to the experimentally measured scattering intensity. This is the goal of multi-objective optimisation, to obtain more realistic estimates by avoiding overfitting experimental data.

\begin{figure}[htb]
\centerline{\includegraphics[width=\textwidth]{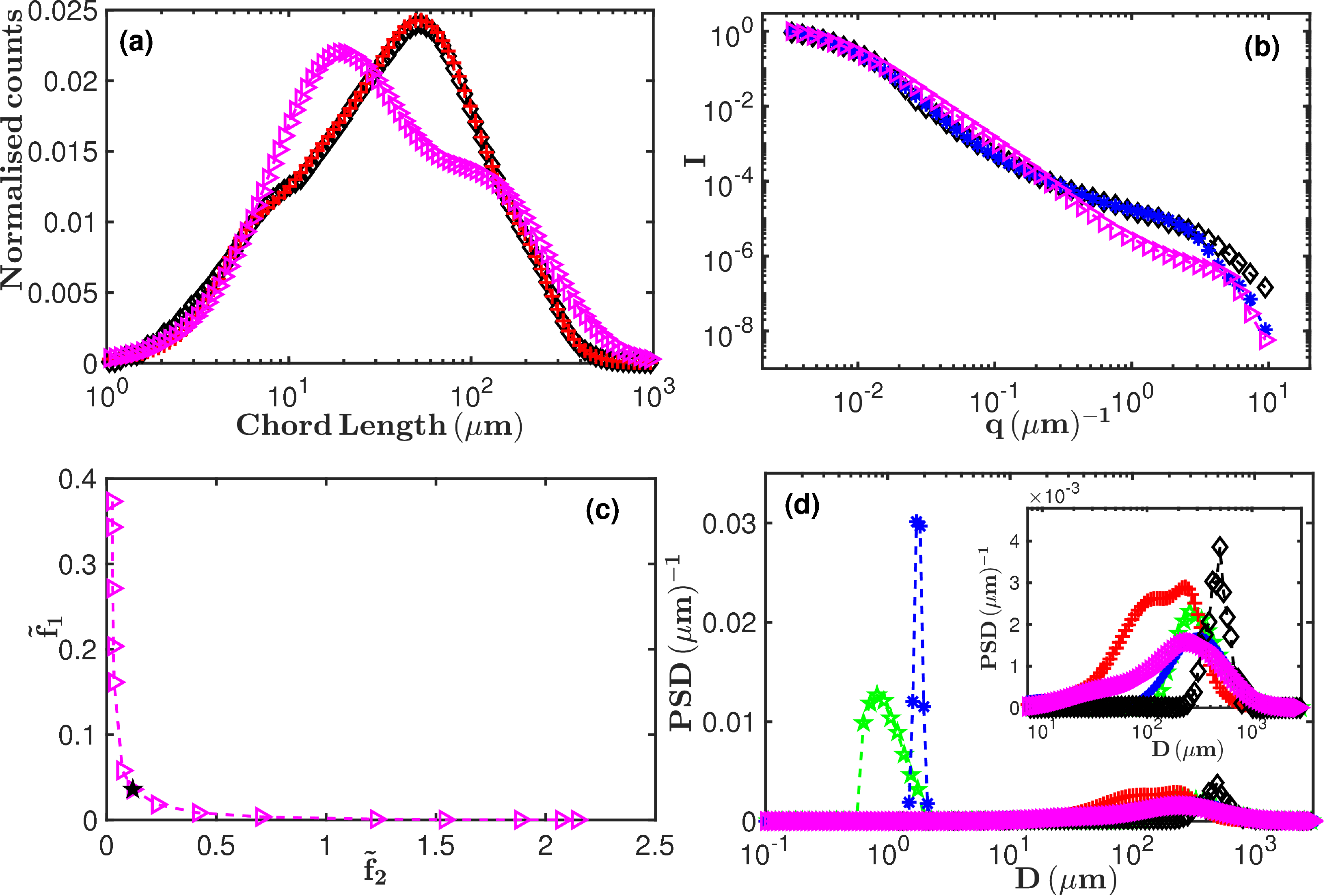}}
\caption{The experimentally measured and estimated CLD, scattering intensity and volume based PSDs for the MET 125-180$\mu$m sample similar to the case of the PS 300-500$\mu$m sample in Fig. \ref{fig1}. The symbols are the same as in  Fig. \ref{fig1}. The inset in (d) is a blow up of the PSDs between $D=10\mu$m and $D=3000\mu$m for clarity.}
\label{fig2}
\end{figure}

The case where both complete elimination and suppression of artificial peaks occur is seen in the MET 125-180$\mu$m sample. Due to the rod-like shape (see supplementary information for sample images) of the particles, the measured scattering intensity (black diamonds in Fig. \ref{fig2}(b)) show a broad shoulder at larger $q$ ($q\gtrsim 1\mu$m$^{-1}$) values, which is not associated with spherical particles of the same size. Hence, in order to fit this measured scattering intensity data with a spherical model, the optimisation solver introduces a particle size mode at $D\approx 1\mu$m in addition to the one at $D\approx300\mu$m. This situation is seen both in the estimated volume based PSD from the minimisation of the objective function $f_2$ in Eq. \eqref{eq3} (blue filled symbols in Fig. \ref{fig2}(d)) and from the Mastersizer (green pentagrams in Fig. \ref{fig2}(d)). The particle size mode at $D\approx 1\mu$m is clearly an artifact judging by the corresponding estimate from the Morphologi instrument, which is shown by the black diamonds in Fig. \ref{fig2}(d). This estimated volume based PSD from the Morphologi instrument suggests there is no particle size mode in the MET 125-180$\mu$m sample at particle sizes around 1$\mu$m.

Similar to the case of the PS 300-500$\mu$m sample, the MET 125-180$\mu$m sample contains particles with bumps on the surface. Hence the measured CLD for this sample contains a small left shoulder as shown by the black diamonds in Fig. \ref{fig2}(a). Hence, the estimated volume based PSD obtained from the minimisation of the objective function $f_1$ in Eq. \eqref{eq2} for CLD, broadens to the left, with a minor peak at $D\approx100\mu$m. This is in addition to the main peak at $D\approx250\mu$m. This is shown by the red crosses in Fig. \ref{fig2}(d). 

The Pareto curve obtained from the minimisation of the objective function $F$ in Eq. \eqref{eq6}, along with the knee point of the curve are shown by the magenta triangles and black filled symbol in Fig. \ref{fig2}(c) respectively. The artificial peak in the estimated volume based PSD at $D\approx 1\mu$m from laser diffraction, and the broadened left shoulder and artificial peak at $D\approx100\mu$m from the corresponding estimate from CLD have either been eliminated or suppressed in the estimated volume based PSD at the knee point of the Pareto curve. The estimated volume based PSD at the knee point of the Pareto curve is shown by the magenta triangles in Fig. \ref{fig2}(d). 

Similar to the case of the PS 300-500$\mu$m sample, the estimated CLD at the knee point of the Pareto curve (magenta triangles in Fig. \ref{fig2}(a)) give a poorer fit to the experimentally measured CLD. The estimated CLD obtained from the minimisation of the objective function $f_1$ (red crosses in Fig. \ref{fig2}(a)) gives a better fit to the experimentally measured CLD. Similarly, the estimated scattering intensity at the knee point of the Pareto curve (magenta triangles in Fig. \ref{fig2}(b)) gives a poorer fit to the experimentally measured scattering intensity (black diamonds in Fig. \ref{fig2}(b)). The estimated scattering intensity obtained from the minimisation of the objective function $f_2$ for laser diffraction (blue filled symbols in Fig. \ref{fig2}(b)) gives a better fit to the experimentally measured scattering intensity.

\section{Conclusions}
\label{sec5}

We have presented an algorithm which is capable of removing or significantly reducing artificial peaks occurring in PSD estimates. The algorithm is based on a fusion of CLD and laser diffraction data. The fusion is done by means of multi-objective optmisation using the weighted sum approach. The algorithm has been applied to CLD and laser diffraction data from two different particle suspensions. The results show the promise held by this multi-objective optimisation approach in obtaining more accurate PSD estimates. 

In situations where an artificial peak is produced, and is significantly separated from the true peak, the algorithm produces a solution where the artificial peak is completely eliminated. In the situation where the artificial peak is too close to the true peak, the algorithm produces a solution in which the artificial peak is significantly reduced.

This algorithm is particularly useful in real-time estimates of PSD from data obtained with inline sensors.
This is because various factors due to the process conditions could lead to PSD estimates with artificial peaks. The occurrence of artificial peaks in PSD estimates could be very misleading in particulate processes where the PSD is a critical attribute of the end product. Crystallisation is a good example of a process in which the PSD of the end product is a critical quality attribute. The development of algorithms, which are capable of obtaining very accurate PSD from multiple data streams will allow the process to be monitored more efficiently and eventually controlled.

\section*{Acknowledgement}

This work was performed within the UK EPSRC funded project \\
(EP/K014250/1) `Intelligent Decision Support and Control Technologies for Continuous Manufacturing and Crystallisation of Pharmaceuticals and Fine Chemicals' (ICT-CMAC). The authors would like to acknowledge financial support from EPSRC, AstraZeneca and GSK. The authors are also grateful for useful discussions with industrial partners from AstraZeneca, GSK, Mettler-Toledo, Perceptive Engineering and Process Systems Enterprise.

 \newpage
 
 \setcounter{section}{0}

 \setcounter{equation}{0}
 
 \setcounter{figure}{0}

 \setcounter{footnote}{0}
 
 \begin{center}
 \LARGE{\textbf{Supplementary Information}}
 \end{center}
  
\section{Sample images}

\begin{figure}[htb]
	\centerline{\includegraphics[width=\textwidth]{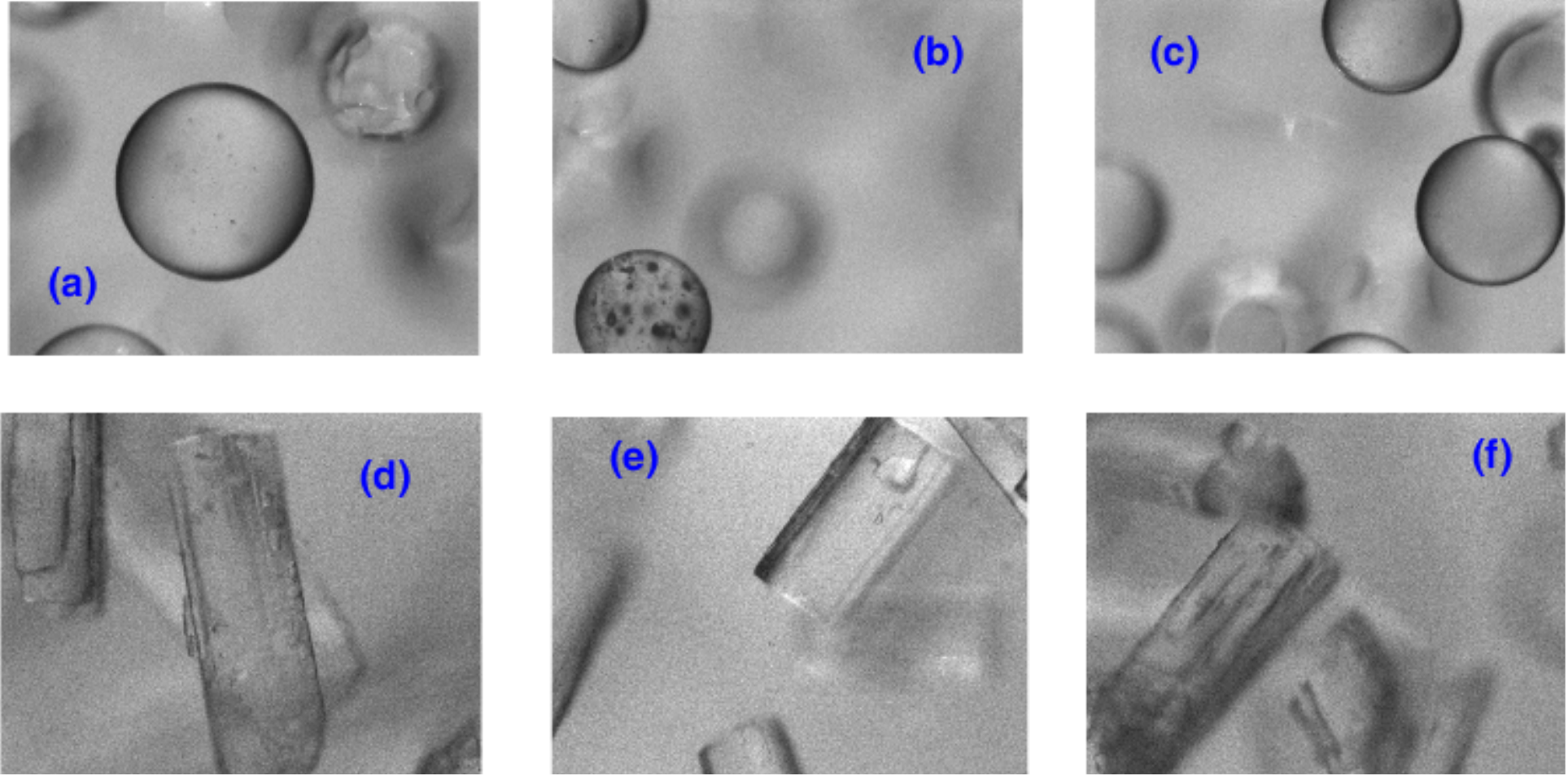}}
	\caption{Sample images for the PS 300-500$\mu$m (a)-(c) and MET 125-180$\mu$m (d)-(f) particles described in the main text are shown. Each image has a width of 1300$\mu$m and a height of 890$\mu$m.}
	\label{figs1}
\end{figure}

Some sample images of the materials analysed in this work are shown in Fig. \ref{figs1}. The images were captured with the Mettler Toledo particle vision and measurement V819 instrument.

\end{document}